\documentclass{aastex631}
\shortauthors{mal et al.}
\graphicspath{{./}{figures/}}

\begin{document}

\title{Redshift periodicity and its significance for Recent observation}

\author{Arindam Mal}
\affiliation{Indian Space Research Organization,Space Applications Centre, Ahmedabad,India}

\author{Sarbani Palit}
\affiliation{ Indian statistical Institute,CVPR unit,Kolkata,India}


\author{Sisir Roy}
\affiliation{National Institute of Advanced Studies, IISC Campus, Bangalore,India}






\begin{abstract}

Recent observational evidence in extra galactic astronomy, the interpretation of the nature of quasar redshift continues to be research interest. Spectrum observation of high redshift quasar is young in nature. Observational evidence discuss on physical interpretation of redshift periodicity with statistical confirmation.
Karlsson observed redshift periodicity at integer multiples of 0.089 in log scale and Burbidge observed redshift periodicity  integer multiple of 0.061 in linear scale  .Data analysis is important in order to form correct interpretations of the observed phenomena. Since Singular value decomposition (SVD) based periodicity  estimation is known to be superior for  noisy data sets, especially  when the data contains multiple harmonics and overtones, mainly irregular in nature, we have chosen it to be our primary tool for analysis of the quasar-galaxy pair redshift data.  Kernel density estimation has been performed for estimating the bin width as proper computation of this quantity is crucial for the correctness of the analysis and prevention of over smoothing of the data.We observed fundamental periodicity to be an  integer multiple of  0.063 and 0.0604 using method1 and method2 in the transformed quasar redshift data with 95\% confidence interval in linear scale.
Our results clearly establish that    redshift is quantized for quasar-galaxy pair data and its histogram exhibits periodic peak(s). At last briefly discussed on physical interpretation of quantized redshift for quasar and galaxy.Hoyle Narlikar theory of gravity explain the Mystery in recent observation.

\end{abstract}



\section{Introduction} \label{sec:intro}

The cosmological hypothes is  defines the total observed redshift as :

\begin{equation}
(1+Z)=(1+Z_c)(1+Z_{Nc})
\end{equation}

where $Z$ is the total observed redshift,$Z_c$  is the redshift due to cosmological contribution and $Z_{NC}$ is the redshift due to non cosmological origin by \cite{Bell1973}.
According to standard theory, quasar redshifts are caused by expansion of the universe. If one plots quasar redshift against apparent brightness, one gets a non- linear relation which implies that linear Hubble relation is not valid for high redshift quasar by \cite{Roy2007,Roy2000}.


   
 H.Arp  observed the physical association of quasar and galaxy (Refs.~\onlinecite{Arp1966}). Interestingly however, it has been observed by Arp that two different objects in the universe  quasi stellar object (QSO) and galaxy are close to each other but have different redshifts. An observational evidence of quasar of Z= 2.114 was found very close to the nucleus of the galaxy NGC7319 with Z= 0.022. In general, Gravitational lensing explains the association of high redshift quasar with galaxy in a few cases but  Arp  has observed  filaments connecting the high redshift quasar with low redshift galaxy; prominent examples being NGC4319 and MK205,NGC 3067 and 3C232 etc.  Some quasars exhibit jets of unknown nature (3C345 in the vicinity of NGC6212), while in some cases, moving structures were found be radio observation along with jets.
 Lopez observed (Refs.~\onlinecite{Lopez2004}) two emission line objects with redshift greater than 0.2 in the optical filament apparently connecting the Seyfert galaxy NGC 7603(Z=0.029),NGC 7603B(Z=0.057) to it's companion,this leads to possible examples of anomalous redshift.Majority of the  findings lead to the conclusion that quasars are ejected from galactic nuclei and redshift is an intrinsic parameter. This naturally raises doubts regarding our current understanding of the significance of the redshift. There are two possible interpretations of physical association; either QSOs with different redshift are objects at different distances, or, non cosmological redshift accounts for QSOs possessing different redshifts at the same distance.

After the discovery of quasars, they were assigned great physical distances because of their high redshifts  (Refs.~\onlinecite{Matthews1963}). However, evidence started to emerge that they were ejected from nearby galaxies and their redshifts were intrinsic (Refs.~\onlinecite{Arp1966}). If purely intrinsic, it would yield a redshift –distance relation of redshift as inverse of age squared (Refs.~\onlinecite{Arp1990}) . Recent geometric tests of the redshift –distance relation (Refs.~\onlinecite{Anderson2012}) too lend support to this theory. Another explanation provided  by (Refs.~\onlinecite{Narlikar1980}),  (Refs.~\onlinecite{Arp1990}) claim  that high redshifts have an intrinsic component which  might be emitting radiation from matter of an younger age. Many researchers
(Refs.~\onlinecite{Narlikar1993, Arp2003, Napier2003, Galianni2005}) argue in favour of quasars ejected from galactic nuclei.

Current theories predict the evolution process, that supermassive black holes begin their lives in the dust-shrouded cores of vigorously star-forming "starburst" galaxies before expelling the surrounding gas and dust and emerging as extremely luminous quasars.Recently observed GNz7q by (Refs.~\onlinecite{Fujimoto2022}), has exactly both aspects of the dusty starburst galaxy and the quasar. This observation  lacks various features that are usually observed in very luminous quasars.The central blackhole of GNz7q is still in a young and less massive phase at high redshift.

Next GN-z11 (Refs.~\onlinecite{Linhua2021}) was photometrically selected as a luminous star-forming galaxy candidate at redshift $z>10$ is known to be "oldest galaxy".The age of GN-Z11 is estimated to be only 70 milion year and moderately massive suggest that this young galaxy was born and grew rapidly, also the fact the evidence of carbon and oxygen in GN-Z11 indicates that this galaxy is not the first (metal- free) galaxy in the universe also direct that it is second generation of  galaxy.The detected light of carbon and oxygen suggest special physical condition not found in present day galaxies.Its accurate redshift remained unclear.

 Also observation of metallicity evolution at high redshift quasar  (Refs.~\onlinecite{Juarez2009}) raise the question on evolution process.Their observation conclude that abundance of carbon relative to silicon and oxygen also does not evolve significantly.
 
\section{Methodology}
\label{sec:methd}

\subsection{Proposed Method}
\label{subsec:methd}

 H.Arp (Refs.~\onlinecite{Arp1994,Arp2001}) observed that a  cluster of high redshift quasars appears to be physically connected with a lower redshift galaxy .  He claimed that their redshifts do not indicate distance and furthermore, redshifts of those clusters are quantized and obey a simple formula:

\begin{equation}
\frac{1+Z_{k+1}}{1+Z_k}=1.227
\end{equation}

where $Z_k$ is the redshift of a quasar and $Z_{k+1}$ is the next higher redshift.
Karlsson (Refs.~\onlinecite{Karlsson1977}) observed that the peaks of the histogram of the redshifts, form a mathematical series $Z=0.061, 0.3, 0.6, 0.96$. Hawkins (Refs.~\onlinecite{Hawkins2002}) found non existence of periodicity, but their methodology was challenged by Napier (Refs.~\onlinecite{Napier2003}).  Next, Tang (Refs.~\onlinecite{Tang2005, Tang2008}) claimed the non- existence of periodicity with a dataset that was 15 times larger than the  previous one.   Duari (Refs.~\onlinecite{Duari1992, Duari1997}) found redshift periodicity statistically. Fulton (Refs.~\onlinecite{Fulton2018}) explains redshift periodicity using ejection velocity computations. Mal (Refs.~\onlinecite{Mal2020}) have already shown the existence of periodicity of redshift for quasar as well as for galaxy; here we have analysed the data for quasar-galaxy pair datasets and try to understand the physics behind the quantization of redshift in such datasets.
 
 We examine the existence of  redshift periodicity in quasar-galaxy pair data following the procedure outlined here. After the initial selection of data in accordance with the flag values provided,  selection of the  bin width for formation of the histogram is optimized as in (Refs.~\onlinecite{Shimazaki2007,  Shimazaki2009}) minimizing the cost function for the overall data set  of Solan Digital Sky Survey’ sample (SDSS DR-7). Next, an SVD- based method  (Refs.~\onlinecite{Kanjilal1995}) has been adapted for estimating the fundamental periodicity present in the histogram of the red shift data.
(Refs.~\onlinecite{Mal2020}) have established the  superiority of  the SVD based approach of periodicity  detection over the periodogram-based approach.  The periodogram is not a suitable tool for data of a quasi-periodic nature or a dataset containing multiple periodic components and a large number of overtones or a somewhat irregular periodic part. Hence, periodicity detection for the quasar-galaxy pair data examined in this article has been performed using the SVD based method.

\subsection{Description of dataset}
\label{subsec:desc}

The SDSS DR7 database contains information pertaining to galaxies with low redshift -quasar pair, where the quasars are projected within 100 Kpc of the galaxy. A total of 97489 galaxy/ quasar pairs are reported from a sample of 105783 spectroscopic quasars and 798948 spectroscopic galaxies. This database contains spectroscopically observed galaxy/quasar projections that can be used to study quasar absorption-line systems arising from known galaxies with redshifts between $0 <Z< 0.6$ and quasar redshifts $0<Z<3.5$.

\subsection{Data selection}
\label{subsec:dat}

The quasar and galaxy redshift data was collected from SDSS DR-7  (Refs.~\onlinecite{Cherinka2011} )and redshifts corresponding to  Ca-II and Na-I flags showing  warning were rejected. Narlikar (Refs.~\onlinecite{Narlikar1993}) explain that if quasars are physically connected with a parent galaxy,  the redshift of each quasar must be transformed to reference frame of the putative parent as :
\begin{equation}
1+Z_0=\frac{1+Z_c}{1+Z_p} \label{eq:putative}
\end{equation}

where $Z_0$ is the transformed quasar redshift, $Z_c$ is the observed companion quasar redshift and $Z_p$ is the observed redshift of the object.
The data was transformed to rest frame as proposed by  (Refs.~\onlinecite{Narlikar1980}) according to
equation (3).
The transformed quasar redshift  data  $Z_0$
with values greater than 0.8 were selected for further processing, in order to avoid the  possibility of mistakenly including  galaxy redshift data.

\subsection{The procedure for analysis}
 As proposed by Mal (Refs.~\onlinecite{Mal2020}), the two main stages of the approach consist of the formation of an appropriate histogram after determining the optimal bin width and application of SVD for periodicity determination. For the convenience, the procedure is briefly outlined here.

{\bf Determination of optimal bin width and histogram formation}: The optimal bin width  is obtained by minimizing the mean integrated square error (Refs.~\onlinecite{Shimazaki2007, Shimazaki2009}) for  the entire data set. optimum binwidth is 0.0029 for the Analysis of transformed quasar redshift data.

{\bf Matrix formulation \& application of SVD}: In order to examine the existence of a periodic component, candidate period lengths are selected and the input matrix for application of SVD, formed accordingly.  For a candidate period length of say, $L$,  the corresponding data matrix $A$ is formed by partitioning the data  into contiguous segments of length each and placing each segment (aligned in phase) as a row of $A$. This matrix is used for singular value decomposition. We follow two approaches outlined below:
Our first method (SVR1) is defined as the ratio of the first two singular values. A plot of this ratio versus row length is called the SVR1 plot which shows the presence of multiple peaks at integer multiples of the value of the fundamental red shift.
The second method (SVR2) for measuring periodicity is defined as the ratio of residual energy to the energy content of the aperiodic signal. The plot of this quantity versus row length is called the SVR2 plot which shows repeated peaks at integral multiples of the value of the fundamental red shift.

\section{Results of simulations}
\label{sec:res}

The fundamental period of the periodic component present is determined as the product of the peak location of the histogram and the optimal bin width.  A peak location is selected as the base period if peaks are also observed at integral multiples of that peak location. The 95\% confidence interval of the base period was determined using resampling of the transformed redshift values $Z_0$ . A Monte Carlo type simulation was conducted over 1000 iterations to compute the confidence interval. The simulations confirm that the reported periodicity of 0.063 has a  95\% confidence interval of  [ 0.0402, 0.0891] using the first  method SVR1 and  0.0604 using  the second method SVR2 in the interval of  [0.0604,  0.0661]. The SVR spectrum exhibits the periodic peak as shown in Figure~1 and Figure~2.
The redshift periodicity of quasar has been observed without transformation in the redshift range ($z>0.03$). The quasar redshift periodicities having the values 0.0906 and 0.1533 has been detected within the 95\% confidence intervals of [0.0744, 0.1114] (using method1) and [0.1259, 0.1811] (using method2). Moreover, the redshift periodicity of galaxies have also been observed in this dataset within the redshift range of ($ 0.03<z<0.1$) of base periodicity 0.0015 and 0.0032 using method1 and method2 corresponding to 95\% confidence interval of [0.0018, 0.0045] and [0.0029, 0.0085], respectively. Apart from that, galaxy redshift periodicity of 0.0032 and 0.0044 has been observed using the method1 and method2 corresponding to 95\% confidence interval of [0.0024, 0.0149] and [0.004, 0.0162] in the range of ($ 0.1<z<0.2$) respectively.

Some salient observations are as follows. Firstly, the existence of at least one periodic component is observed for several ranges of redshifts of quasars and galaxies, secondly observation may be made regarding the presence of additional periodic components which are  noticed in Fig.1 (identified as red and pink vertical marker).This is confirmed by a secondary strong peak and a peak occurring at a location, which is a multiple of the location of the secondary peak 0.077 identified in pink colour, whereas primary peak at integer multiple of  0.0603 identified in Red colour.
The redshift periodicity of quasars and galaxies may indicate  that evolution of quasars into galaxies have occurred over time.

    \begin{figure}[h!]
    \centering
    \includegraphics[width=9cm,height=6cm]{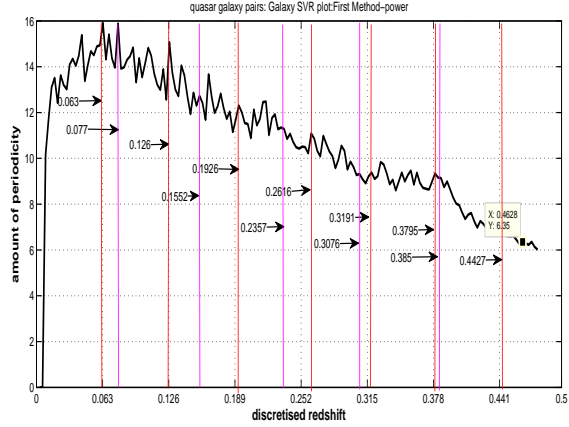}
    \caption{Quasar Galaxy pairs SVR Spectrum :  SVR1  }
    \label{fig1:app1}

    \end{figure}

    \begin{figure}[h!]
    \centering
    \includegraphics[width=9cm,height=6cm]{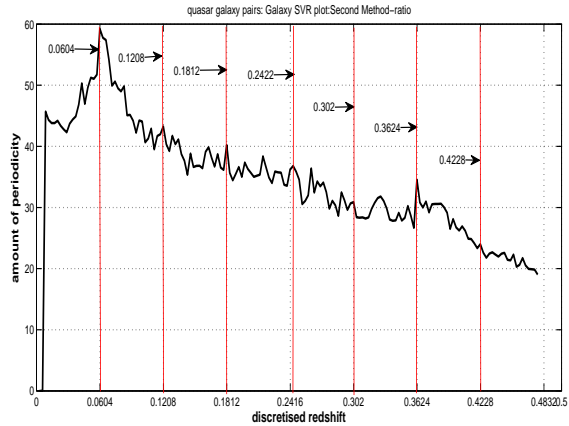}
    \caption{Quasar Galaxy pairs SVR Spectrum : SVR2}
    \label{fig1:app2}

    \end{figure}


\section{Physical Interpretation and Discussion: }
\label{sec:interpretation}
\subsection{Interpretation of the result}
Physical interpretation of redshift periodicity confirm  the objects which have same redshift are physical members of same clusters, or evolution of quasar is concentrated about epochs spaced in this way, i.e it must have some sort of "periodic "structure / " crystalline" structure or  It can be interpreted that non cosmological origin is more stronger than other components of redshift as discussed in equation 1.

\subsection{Discussion}
There are two conventional alternatives to account for the discretization of redshift or anomalous redshift of QSO, one is Doppler shift and another one is gravitational redshift, but they cannot explain physical association of low redshift galaxy with multiple QSOs and evolution process.


The extremely active field of gravitational lensing has also made rapid strides in locating and imaging of a variety of lensed quasars at various wavelengths. High resolution radio imaging was used by Kratzer (Refs.~\onlinecite{Kratzer2011}) to resolve a large separation lensed quasar at $Z=2.197$. Zimmer (Refs.~\onlinecite{Zimmer2010}) used archival Chandra data to construct an x-ray light curve of all four images of the quadruply lensed quasar at Z=1.695. High energy gamma rays were used for gravitational lensing for the first time by Barnacka (Refs.~\onlinecite{Barnacka2011}). Corredoira (Refs.~\onlinecite{Corredoira2006}) propose weak gravitational lensing by dark matter as the cause of the statistical correlation between low and high redshift object. However,Scranton (Refs.~\onlinecite{Scranton2005})  have contradicted them and claimed that the correlation found was an ad-hoc fit of the halo distribution function to an angular cross correlation with very small amplitude of the galaxy selected photometrically. Even to this day, standard cosmology is unable to provide an explanation for the correlation of galaxies and QSOs. Gravitational lensing explains the association of a high redshift galaxy with a single quasar using the standard cosmological model but fails to explain multiple gravitational lensing within the galaxy.

The generally accepted idea is that a galaxy is surrounded by background QSOs. However the question of statistical analysis of the background /foreground object probability in a small area distributed according to its position and average density in any line of sight, raises much controversy in the astrophysical community. The main approach here is to demonstrate that two extragalactic objects with very different redshifts  may be physically neighbors in reality. The physical association of quasar and galaxy is an example of the two cosmological objects being neighbors of each other.

Physical association of quasar and galaxy also explain using super massive black hole(SMBH) and Active galactic Nuclei(AGN). SMBH can grow in size up to billions of solar masses from “seed”, the grow process depend on feeding mechanism from surrounding gas, Astronomer think that every galaxy have extremely bright centre region called Active galactic Nuclei(AGN) and think that they are powered by SMBH in their centre. The most luminous of all the AGN are quasars. Recent discovery of J0313-1806 discovery leads to a question of formation of it. J0313-1806 is thought to sit inside a galaxy with a very active region, the host galaxy produces 200 solar masses worth of star per year. The SMBH that powers this quasar formed just 670 million year after Big-Bang, which ask question on how “seed” grew from to form and how this black hole stellar !. Next researchers propose alternative mechanism called ‘direct collapse’, but truth is unknown.

One of the outstanding mysteries in astronomy today is: How did supermassive black holes, weighing millions to billions of times the mass of the Sun, get to be so huge so fast at the edge of universe.

Recent observations of extra-galactic objects that do not appear to be consistent with the cosmological hypothesis that their redshifts arise from the expansion of the universe.

Our statistical analysis of redshift data clearly establishes the existence of discrete redshift using quasar galaxy/ pair red shift dataset. In the view of the above difficulties with the conventional model, we look for alternative models which can explain the periodicity of redshifts. One such model is proposed as by Hoyle-Narlikar (Refs.~\onlinecite{Narlikar1980,Hoyle1964,Hoyle1966}) which tries to explain the periodicity of redshifts based on Variable Mass Hypothesis(VMH). According to this, the inertia of matter arises due to interaction of other matter in the universe, the matter created from zero mass surface based on quantum principle.
The excess redshift does not Aries from high speed of ejection but from the low mass of the newly created matter.
Narlikar (Refs.~\onlinecite{Narlikar1980})explain that the ejected QSO can be bound to the parent galaxy with typical separation of the order of ~100-200kpc.Hoyle (Refs.~\onlinecite{Hoyle2000}) explain redshift periodicity using variable mass hypotheses. Redshift periodicity is a direct confirmation of the fact that quantization in redshift implies quantization in mass.VMH theory predict that the age of an object usually measured from zero mass epoch on its world line hence higher the redshift of quasar younger it is, this is well matched with recent observation of extragalactic objects: HD1,HD2,GNz7Q,GNZ11.

Another model proposed and elaborated upon by  Roy et al. (Refs.~\onlinecite{Roy2007,Roy2000}) based  on Wolf mechanism is known as Dynamic Multiple Scattering (DMS) theory. It is shown in a recent paper that DMS can explain the redshift for Galaxy-Quasar association. This depends on the property of the environments around galaxies as well as that of quasars. This mechanism mimics Doppler shift even in the absence of relative motion of the observer and the source.

Wolf(Refs.~\onlinecite{Wolf1986}) explains correlation induced spectral shift as well as the broadening and shifts of the spectral line.  Roy et al.(Refs.~\onlinecite{Roy2007,Roy2000}) analyzed statistically the Veron-Cetty (V-C) quasar catalogue (2006) and SDSS DR3 data set and concluded that the Hubble law is linear up to small redshifts less than 0.3($z<0.3$) but nonlinear for higher redshift, which adds fuel to the cosmological debate. Roy et al (Refs.~\onlinecite{Roy2000}) explain the broadening of the spectral line using dynamic multiple scattering. They also found a critical source frequency below which no spectrum can be observed for a particular medium. The broadening due to multiple scattering is more than the shift due to cosmological effect. In our future works we will discuss the possible explanation of periodicity of redshifts within DMS framework.

There are two fundamental objection against the ejection hypotheses. First, in the ejection model, ejection is always away from us means redshifted. second, The apparent velocity difference is a large fraction of the speed of light.Many researchers has explained the 1st problem using different selection mechanism, also this is well explained using wolf effect based method . The spectral shift of quasar depend on the surrounding scattering medium of quasar and ejection angle.
Peter.M(Refs.~\onlinecite{peterM2006}) explain  apparent redshift  components of quasar-galaxy association using parametric model in the wolf framework.
The second problem does not exist in the relativistic slingshot process by Saslaw(Refs.~\onlinecite{saslaw1974})of ejecting black-hole. 


\section{Conclusions and future work}
\label{sec:conclusion}

   We have observed the existence of  redshift periodicity in quasar / galaxy pair redshift data without transformation, also the existence of periodicity in this data can be clearly perceived upon analyzing it after transformation to rest frame. Since data binning has been performed using kernel density estimation method and not in a heuristic fashion, our results directly contradict the proposition reported by some papers that redshift periodicity observed is actually due to selection effect of the data binning by Basu (Refs.~\onlinecite{Basu1978}). other kind of noise in the data may be influence to redshift periodicity due to that fundamental periodicity may be within the band. The most significant contribution of the present work is the analysis of the paired redshift data and its periodicity detection and estimation using an SVD based approach rather than the conventional Periodogram/FFT based approaches. The applicability of this approach, also used in Mal et al.(Refs.~\onlinecite{Mal2020}) is further vindicated by the present results. It thus proves to be invaluable for astronomical data analysis since it is able to reveal hidden periodicities better than traditional approaches hence leading to correct interpretation of the data. Fulton (Refs.~\onlinecite{Fulton2018}) has proved the existence of Karlsson peaks using the ejection velocity constraint but it may be noted that the proposed work gives a direct way of estimation of redshift periodicity.This article confirms the existence of redshift quantization in quasar-galaxy associated dataset and supports the Hoyle-Narlikar model (Refs.~\onlinecite{Hoyle1964, Hoyle1966}) and Narlikar–Das cosmological model (Refs.~ \onlinecite{Narlikar1980}) and (Refs.~\onlinecite{Hoyle2000}) that quasars are  ejected from galactic nuclei and explain the process of evolution. 
   
   Recent observational result should be mentioned that most distance quasar are most metal deficient, but surprinsigly high metal abundances were found in high redshift quasar. The question are how metal produced by rapidly evolved stellar population around quasar, so soon after the bigbang.The only way to produce heavy elements are known nuclear process at large stage of their evolution.  Satndard theory can not explain the above observation, if redshift is not intrinsic parameter.
   
   Also if quasar are at cosmological distances, the velocities of theses moving structure should be super luminous on the otherhand if it is local origin the " super luminous" velocities will be reduced below the velocity of light.
   
   Presently cosmologists are greatly interested in understanding the source of Ly$\alpha$ forest and environment surrounding the galaxy. In order to answer the question of how the forest arises, the variable mass hypotheses explains that lower redshift quasar implies larger and older mass whereas a higher redshift quasar is younger and possesses smaller mass.



   In contrast to the standard model, the absorption feature in Ly$\alpha$ forest should exhibit peaks at multiples of base periodic component (Refs.~\onlinecite{Karlsson1971, Karlsson1973, Karlsson1990}) and (Refs.~\onlinecite{Burbidge1968}).This paper also confirm the periodicity of redshift of quasar and galaxy .  Further experimentation is required for detail understanding of quasar ejected from galactic nuclei using 3D imaging.  

\begin{acknowledgments}
The authors express their thanks to Prof. J.V.Narlikar. of IUCCA for  excellent guidance and suggestions provided during the research time.Authors wants to express their special thanks to Christopher C. Fulton and Martín López Corredoira for providing suggestion during discussion.The authors would like to express their gratitude to Dr. Debiprasad Duari for  his interest in this work.
Authors also acknowledge Prof. Debasis Sengupta from ISI, Kolkata for his help in designing simulations and Ms. Susmita Nandi for her assistance in carrying out some simulations. They would also like to acknowledge H.S.Ravindra of LEOS, Tapan Misra and Atul Shukla  of SAC, ISRO for their encouragement.
\end{acknowledgments}


 \bibliography{arindam_letters_bib}{}                
\bibliographystyle{aasjournal}



\end{document}